\pdfoutput=1

\documentclass[aps,showpacs,preprintnumbers,amsmath,amssymb,
eqsecnum, twocolumn, tightenlines
]{revtex4}

\usepackage{graphicx}

\newcommand{\be}{\begin{eqnarray}}
\newcommand{\ee}{\end{eqnarray}}

 \newcommand{\gsim}{\mathrel{\hbox{\rlap{\lower.55ex \hbox {$\sim$}}
                   \kern-.3em \raise.4ex \hbox{$>$}}}}
\newcommand{\lsim}{\mathrel{\hbox{\rlap{\lower.55ex \hbox {$\sim$}}
                   \kern-.3em \raise.4ex \hbox{$<$}}}}

\newcommand{\ba}{\begin{eqnarray}}
\newcommand{\ea}{\end{eqnarray}}

\begin{document}


\title{ Shocks in Quark-Gluon Plasmas}

\author { Edward Shuryak }
\address { Department of Physics and Astronomy, State University of New York,
Stony Brook, NY 11794}
\date{\today}

\begin{abstract}
Hydrodynamics is known to describe matter created in high energy heavy ion collisions well.
Large deposition of energy by passing jets should create not only the sound waves,
already discussed in literature, but also the shocks waves of finite amplitude. 
This paper is an introduction to relativistic shocks, which go through elementary
energy and momentum continuity argument, to weak shocks treated in Navier-Stokes
approximation, to out-of-equilibrium setting of AdS/CFT. While we have not yet found
numerical solution to corresponding Einstein equations, we have found  a variational approximation
to the sum of their squares. Our general conclusion is that deviations from LS and NS hydrodynamical
shock profiles are surprisingly small, even for strong shocks.
We end with a list of open questions which the exact solution
should be able to answer. 
\end{abstract}
\maketitle

\section{Introduction}
  Shocks are rather well known hydrodynamical phenomenon described in
 the textbooks such as \cite{LL}. 
 Their discussion  in the framework of ultrarelativistic 
 collision of nuclei 
has been initiated by Landau and Belenky  \cite{LandauBelenky} who had  applied it to
matter compression  at the initial time. Important point made in their paper was that very strong shocks (of very large amplitude)
 do not depend on the EOS of the matter $before$ the shock. 

  In 1970's, with the first experiments with ultrarelativistic nuclei at BEVALAC, it has been suggested
  to look for shock waves and Mach cones in excited nuclear matter. Unfortunately hydrodynamics itself has not worked out in this application because the nuclear matter is not a good enough liquid:
  the  nucleon mean free path is comparable to the nuclear sizes.   
Another applications of the shock theory to heavy ion collisions has been considerations of the deflagration and detonation
shocks, propagating from the system's edge inward \cite{deflagration}. Small velocity of such fronts were predicted to induce the so called ``burning log" scenario for RHIC collisions. In practice however
it was later found that a  very long time needed for those shock propagation  is not really available in  real collisions, in which the matter outward gradients are large enough to
 generate strong  radial flow overwhelming the  deflagration.

  Unlike excited nuclear matter, Quark-Gluon Plasma (QGP) turned out to be a very good liquid,
  with the effective mean free path about an order of magnitude smaller than that for nucleons
  in nuclear matter.
  Soon after beginning of the RHIC era of experiments it became
  clear that already an ideal relativistic hydrodynamics
  describes those very well \cite{Teaney:2001av,Huovinen:2001cy,Hirano:2002ds} while the viscous corrections are indeed small \cite{Teaney:2003kp,Romatschke:2007mq}.
  Further recent confirmation came from the next-order applications, such as sound perturbations from initial state
  fluctuations. The calculated   higher angular harmonics of the correlation functions are in good agreement with
the RHIC/LHC data,  see e.g. \cite{Staig:2011as}.
  
    AdS/CFT correspondence has been used to explain the rapid onset of hydrodynamical regime. 
  It started with the  small viscosity-to-entropy ratio predicted in \cite{Policastro:2001yc}
  \be \eta/s=1/4\pi  \label{eqn_son} \ee  
  for any
  strongly coupled theory with the holographic dual. Later those studies has been followed by studies of various out-of-equilibrium 
  settings \cite{Lin:2008rw,Chesler:2010bi,janik2}. Basically all of them found rapid onset of viscous hydrodynamics,
  basically  in the so called ``infalling" time needed to approach the forming black hole horizon.
  
  Unfortunately, the out-of-equilibrium settings just mentioned are all time-dependent, and (except for the first one)
  require solution of the 2+1 dimensional dynamical Einstein equations, which is technically very challenging. 
  In this paper we propose to study the shock waves as an alternative {\em out-of-equilibrium stationary} setting.
  In this problem the Einstein equations
  are basically elliptic and thus allow for much easier treatment than the dynamical (hyperbolic) ones. In fact, we will be arguing below that
  one can use rather simplistic  variational approach and get sufficiently accurate approximation,
  even for strong shocks. (Weak shocks in AdS/CFT setting has already been discussed in Refs \cite{1004.3803,1105.1355} and we will not study/review this case.)
  
  Another tool to be discussed is the so called {\em resumed hydrodynamics} \cite{Lublinsky:2009kv} which suggests
  an approximate way to include higher gradients in sQGP. We will discuss the 
  convergence of the corresponding series in gradients, as well as the corresponding Lublinsky-Shuryak (LS) 
  resumption of those. 
  
Last topic in the introduction is phenomenological: it deals with some estimates for shock which appear due to jet quenching at LHC. We will argue that the energy deposition is large enough to create strong shocks at early time.
The main point is the observation  of very asymmetric events, in which the trigger jet has 
the transverse energy $E_T$ significantly larger than that of the associate jet $E_A$.
Energy of the jet is  deposited into QGP can be as large as 
\be \Delta E\approx E_T-E_A \sim 100 \, GeV \ee
(The first equality we write as approximate since the trigger also looses some energy,
as well as picking up some from fluctuations and trigger bias effect, both of the scale 10\, GeV or so.)

Let us now compare to the volume in which this energy is deposited.
Basic unit of length we use is the holographic  distance to the horizon $1/(\pi T)\approx .15 \, fm$.
The strong shock example we used above produces a width-at-half maximum of about 4 such units,
making it about $\delta r \approx 0.6 fm$ for $T=.4 GeV=1/(0.5\, fm)$ . 
Energy of the jet is  deposited into a cylinder of length $L$ and radius given by the shock width. Its volume is
\be V=L \pi r^2 \sim (10 fm)*3.14*(0.36 \, fm^2 ) \approx 10 fm^3 \ee
If this energy $\Delta E$ is, say, 100 GeV, the deposited energy density $\delta \epsilon \approx 10 \,\,GeV/fm^3$, comparable to 
the QGP pressure in the bulk 
 \be p_{QGP}\approx 4 T^4 & =& 4 *(0.4\, GeV) /(0.5\, fm)^3 \nonumber \\
\sim  \,\, 10 \,\,GeV/fm^3 \ee  
The conclusions is obvious: such jet quenching event changes the stress tensor by some
factor 2 or so. This 
should create what we called {\em strong shocks}, which remain so at least
for some  time. What it means is that the Mach cones from them should be different from those
considered in the literature \cite{CasalderreySolana:2004qm,Chesler:2007sv}: those were assumed to be sounds 
and move with the speed of sound. 
The difference can be seen from the results of the nest section.
 The detailed study of such modified Mach cones can be given elsewhere.

\section{Two continuity equations}
For self-consistency of the paper we briefly review some  well know material.

   Before we do so, let us introduce kinematical notations.
  The shock problem can be considered in two
different frames. In the relativistic notations we will consistently use rapidity variables, in which
Lorentz transformation between them is simply additive: we think it is pedagogically simpler than using 
Lorentzian square roots.

Frame A: the matter in front of the shock is
considered to be at rest (zero rapidity).
We denote by $Y$ the rapidity of the matter behind the shock, and by $y$ the rapidity of the shock itself.

Frame B: The shock is at rest and 
 the shock front is stationary.  Shifting all rapidities by $-y$, one get inflowing matter with the rapidity $Y-y$ and for 
 out floating matter the rapidity is $y$.
 
One difference  of the QGP with the usual non-relativistic liquids  is that
the conservation of matter -- of the number of atoms or at least some conserved
vector current -- does not play an important role. Indeed
the conservation of two vector currents available -- that of the baryon number and strangeness 
 -- can well be ignored, as the values in question are very small. 
 Therefore, the matter
will be described  by its stress tensor alone.
The fundamental laws to be used are those of the energy and momentum conservation. 
 
There are no time derivatives and the vanishing divergences in hydro equations integrate
to the continuity equations, of the $T^{01}$ and  $T^{11}$ :
\be 4 p_f cosh(y-Y) sinh(y-Y)   \nonumber \\ = 4 p_i cosh(y) sinh(y)  \\
3 p_f sinh(Y-y)^2+p_f cosh(Y-y)^2 \nonumber \\ = 3 p_i sinh(y)^2+p_i cosh(y)^2
\ee
where we have used standard relativistic local expression for the stress tensor, with $e_{i,f},p_{i,f}$ being the initial/final energy density and pressure on both sides
of the shock. 

\begin{figure}[!t]
\begin{center}
\includegraphics[width=6.5 cm]{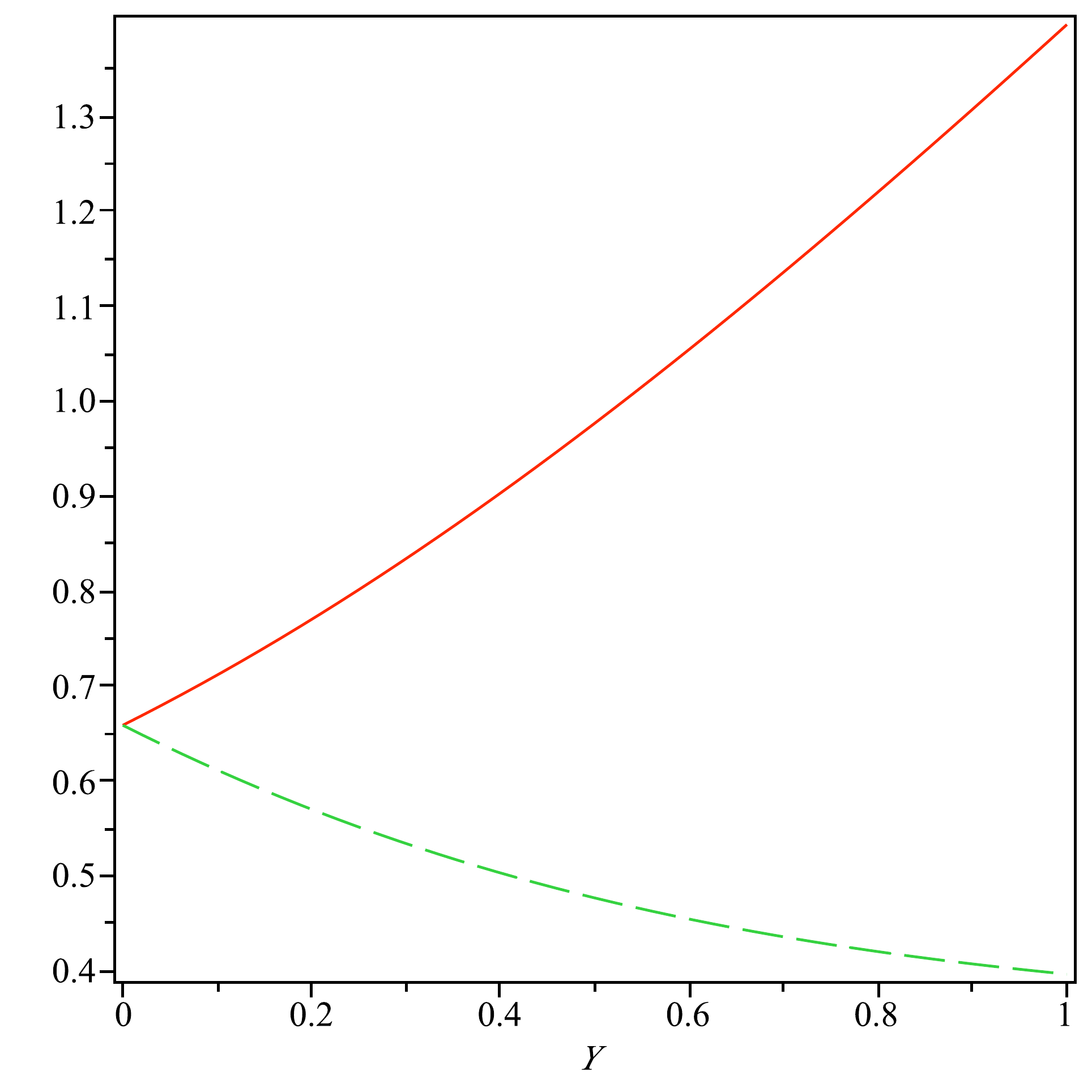}
\includegraphics[width=6.5 cm]{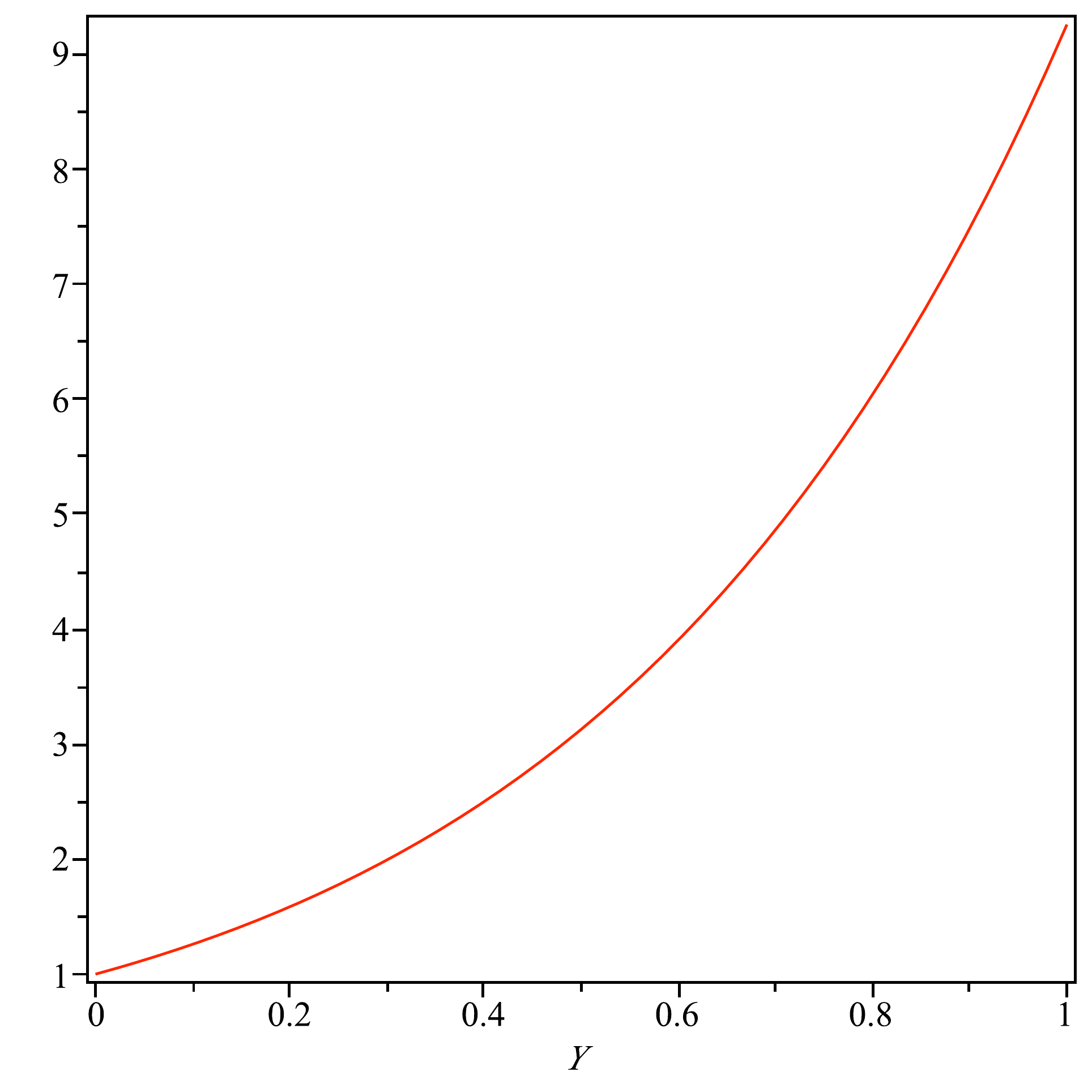}
\end{center}
\vspace{-5ex}\caption{(color online)
(Upper plot) The rapidity of the shock $y$ versus the matter rapidity $Y$ is shown by upper (red) solid line,
while the lower (green) dashed line is $y-Y$. (Lower plot) The compression ratio versus $Y$.   }
\label{fig_compression}
\end{figure}

Let us start with the simplest conformal equation of state,  the same QGP phase on both sides, which means that one can use $e_i=3p_i$. 
Now these  two equations can be written as two expressions for the compression ratio 
\be
{p_f \over p_i}= { cosh(y)sinh(y)   \over cosh(y-Y) sinh(y-Y) } \nonumber \\ = { (3 sinh(y)^2+cosh(y)^2)  \over (3 sinh(-Y+y)^2+cosh(-Y+y)^2)}
\ee
The second equality is in fact 
the equation for $y$, provided the rapidity jump  $Y$ of the matter is given. 
This equation has the following  solution
\be y(Y)= Y-(1/2) ln(e^{2 Y}+1-\sqrt{e^{4 Y}+e^{2 Y}+1)}\ee
which is plotted, together with $y(Y)-Y$, in the upper part of Fig.\ref{fig_compression}.

The constant value of the shock rapidity at small compression (compression ratio close to 1, or {\em weak shocks} at $Y \ll 1$) is 
\be y(Y=0)= (1/2) ln(2-\sqrt{3}) \\ \nonumber tanh(y(Y=0))={1 \over \sqrt{3}} \label{sound_rapidity} \ee
It is nothing else but the sound rapidity (velocity). Thus two branches of the solutions are often called the
supersonic and subsonic matter flows.

 As the compression ratio grows,  together with $Y$, to values of the order one $Y\sim O(1)$ 
 one finds the so called {\em strong shocks}, to be discussed in what follows. 
 Note that for initial state perturbations  in heavy ion collisions the compressions are indeed generally of this category.
 When the compression ratio and rapidity jump gets large, those  shocks can be called {\em very strong}. 
 We will not discuss those: and just note that these two quantities  grow very differently because of hyperbolic functions involved.
For example large compression ratio of 10 corresponds to the rapidity jump which is still $Y\approx 1.$.



This has consequences for the ``Mach angle" as the $cos\theta_M= v_{shock}/c$ is approaching one. It is also important for the 
size of the ``shock horizon" which is the distance the shock can travel till freezeout.


\section{Weak relativistic shocks }
Weak, we repeat, means  small rapidity jump through the shock front
\be Y=y_f-y_i\ll 1\ee
In this case, as we will soon see, the gradients are small and this 
classic textbook problem  is  solved in the Navier-Stokes (NS) approximation
(see e.g. \cite{LL}, chapter 87). The principal steps in the relativistic case  are the same
as in non relativistic case, minor modifications are due to (i)
the relativistic kinematics of the flow; (ii) the absence of the conserved matter current, and (iii) different
EOS compared to the nonrelativistic gases usually considered.

As we already stated, we ignore vector current and their conservations altogether,
focussing on the stress tensor. Its relativistic dissipative
part can be written as follows
 \ba \delta T_{\mu\nu}&=&\eta(\nabla_\mu u_\nu + \nabla_\nu u_\mu
 -{2\over 3}\Delta_{\mu\nu}\nabla_\rho u_\rho)\nonumber \\
&&+ \xi(\Delta_{\mu\nu}\nabla_\rho u_\rho) 
  \ea
%
  where the  coefficients  $\eta,\xi$         are  the shear
 and  the  bulk  viscosities.
    the   projection
 operator onto the matter rest frame is
  \be \nabla_\mu\equiv\Delta_{\mu\nu}\partial_\nu, \,\,\,
  \Delta_{\mu\nu}\equiv g_{\mu\nu}-u_\mu u_\nu \ee
  
  The 11 and 01 NS equations, for
   the simplest conformal EOS $\epsilon=3p$, read
  \be  
  p (4u_1^2+1) +\eta {4\over 3} \partial_x u_1 =C_{11}
 \\
 p 4 u_0 u_1 +\eta  \partial_x u_0=C{10}
  \ee
where two  constants in the r.h.s.  can be inferred e.g. from flow (fluxes) far before
the shock $x\rightarrow - \infty$ where the matter is homogeneous and the gradient terms are absent.
Writing the functions as initial values plus modifications
$p(x)=p_i+\delta p(x) , \,\,\, y(x)=y_i+\delta y(x) $
and substituting them to the two equations above, 
one can perform expansion in small terms up to the $second$ order.
 Note that 
the viscosity term can be kept as constant as the gradient is already of the desired
magnitude of smallness. Since $\delta p$ only appears linearly, one can find it from one equation
and substitute it into another, obtaining  thus a closed differential equation for $\delta y(x)$ alone.
It is quadratic in rapidity perturbation and can be rewritten in the following transparent form
\be
 (\delta y)(\delta y- Y) +(\Delta x) {d\delta y \over dx}=0  \label{eqn_weak}
\ee
where, we remind, $Y=y_f-y_i$. The value of the coefficient is ($c_i=cosh(y_i)$)
\be \Delta x = {\eta \over 12 p_i}  {c_i (c_i^2-1)(4c_i^2+9) \over -2 c_i^4+6c_i^2-3} \ee
in which the initial rapidity can be approximated by that of the sound. 

Indeed,
 by construction,  one root of the quadratic form is zero and the second root of the l.h.s. (other than $y_i$) must be
the jump to the final rapidity $y_f$. This is the same generic equation as one gets (for pressure) of  the
non relativistic shock, and its solution is predictably the ``Fermi step function"
\be \delta y(x) = {Y \over 1+exp(-Y x/\Delta x)} \ee
The {\em width of the shock}  for weak shocks
  is parametrically large, $O(1/Y)$, which justifies the applicability of the gradient expansion
 and explains why  it propagates with the speed of sound. 

\begin{figure}[!t]
\begin{center}
\includegraphics[width=7.5cm]{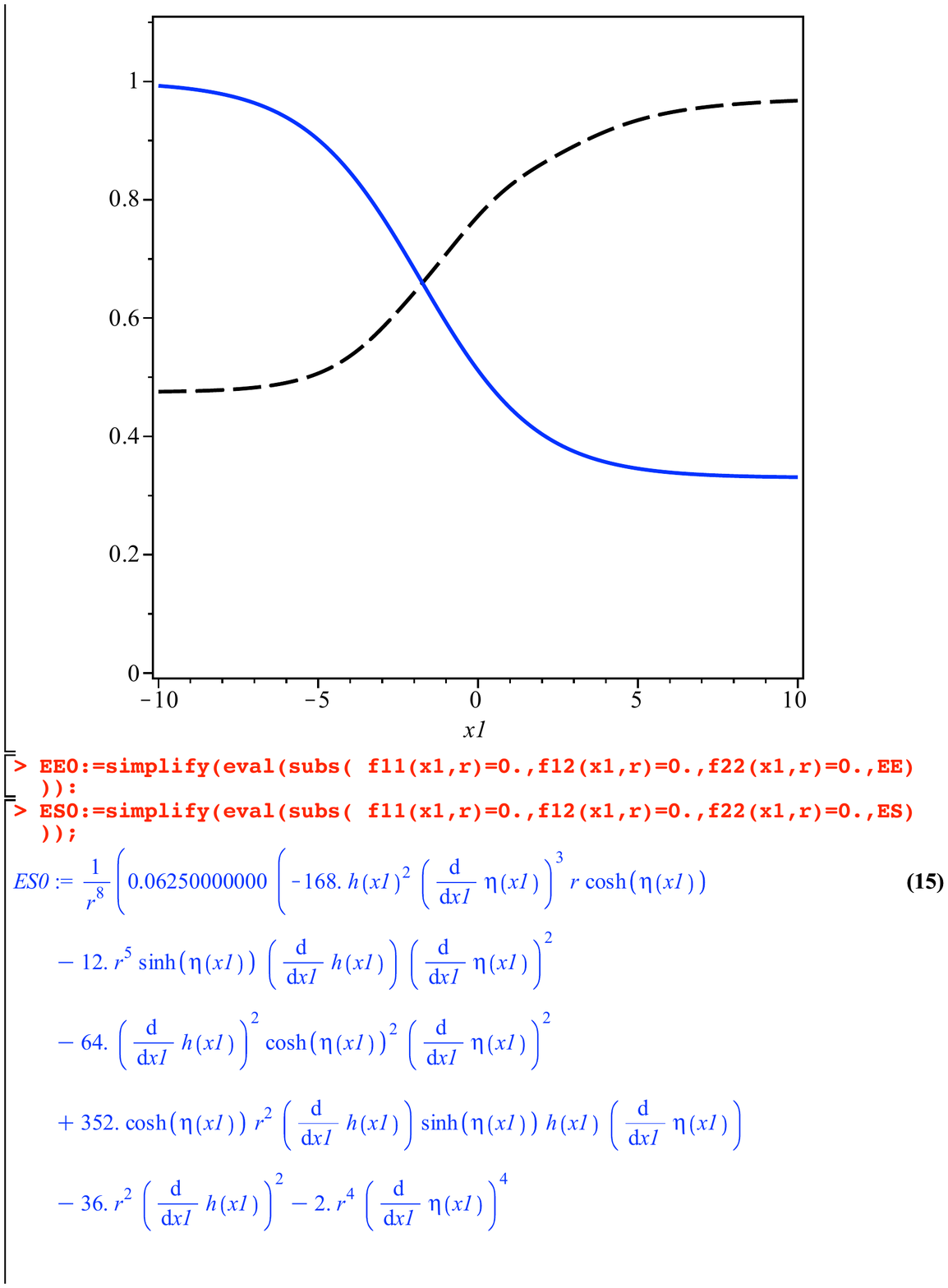}
\end{center}
\vspace{-5ex}\caption{(color online)
An example of a solution to the NS  hydrodynamics, the rapidity (black dashed line) and pressure (blue solid line. Pressure is in units of  its initial value: thus the curve starts near 1 on the left.  }
\label{fig_NS}
\end{figure}

\section{Strong shocks and the resumed  hydrodynamics}
Let us now proceed to  shocks 
 with the rapidity jump $Y=O(1)$, for which there is no apparent small parameter and
 thus the NS equation cannot be expanded. 
 Since pressure appears only linearly, one can easily manipulate $01$ and $11$ equations
 into one single differential equation for rapidity (the second equality)
 \be 
 p&=& {-(4/3)\eta c (dy/dx) +p_i (4 s_i^2+1) \over 4s^2+1} \nonumber \\
&&  = {-\eta s (dy/dx) +4p_i s_i c_i \over 4 s c}
\label{NS_nonexpanded}
 \ee 
 where we use the short-hand notations $$c=cosh(y(x)),s=sinh(y(x))$$ All quantities with the index $i=initial$ are the corresponding values before the shock, at 
 $x=-\infty$. 
The solution to the NS equation cannot be obtained analytically, so we use a numerical solver.
A particular example is shown in Fig.\ref{fig_NS}, for the rapidity and the pressure.
Its parameter were selected at random: the rapidity    jumps up by a factor of 2 and the pressure
down by about factor 3. 

Now, how reliable is the NS solution, for such a strong shock? 
The implicit assumption is that all nonlocal (higher gradient) terms are small. We need to know \\(i) the individual coefficients -- higher viscosities -- of those terms; \\(ii) 
the values of the higher  gradients; \\
and (iii)  the combined effect of their $sum$, or convergence of the series.

For usual fluids  such as water to air, the gradients are usually so small that we do not need higher viscosities. 
In practice  their empirical  values are not even known, and so it would be hard
 to even estimate the magnitude of  those terms. Yet for sQGP we have the AdS/CFT correspondence, which provides in principle a complete set of such higher viscosities, and many of those has already been evaluated in literature.  
The summary of these results and their resummation was the subject of the 
 ``improved hydrodynamics" by Lublinsky and myself
\cite{Lublinsky:2009kv},
LS for short. 

  The linearized correlators in AdS/CFT can in principle provide the values of
all kinetic coefficients, as the coefficient of certain powers of $\omega$ and $k$
 (the frequency and the wave vector). Since there are several kinematically different channels excited by  stress tensor,
these calculations  also provide  sufficient number of crosschecks,
For example, ensuring that the first viscosity always is (\ref{eqn_son}) anywhere viscosity appears.
The ``improved hydrodynamics"
tries to combine all known coefficient into some ``resumed" model functions.
The main result was the 3-term Model 1 which is based on PADE
approximation, which may be called LS1.
 It reproduces exactly eight first coefficients
and overall behavior of the correlators quite well. Furthermore, we found that the second and the third
poles largely cancel each other, and for 
an estimate of  the effect of higher gradients 
we will use our simpler model  LS2 in which the effective viscosity is
\be  
\eta_{LS2}={ \eta_0 \over 1- \eta_{2,0} k^2/(2\pi T)^2 -i \omega \eta_{0,1}/(2\pi T)} 
\ee
written in units in which all eta coefficients are dimensionless numbers. Furthermore, as there is no time dependence in stationary problems
we now discuss, we will only need
one simple coefficient
\be \eta_{2,0} =-{1\over 2} \ee
Due to its sign,  the resumed factor $reduces$ the effect of the NS term as $k$ grows. 
This is indeed what was known for the lowest quasi normal modes (sounds): its imaginary
part grows as $k^2$ till some value, and then it stops growing.

As the rest of this paper below will be discussing the AdS/CFT setting, it is convenient to 
us   the holographic thermal horizon as   the natural unit of length.  Thus
 the horizon location in the holographic coordinate is \be z_h=1/\pi T=1 \ee 
Using it and the holographic value of the first viscosity (\ref{eqn_son}) one finds conveniently
that the combinations which often appears is greatly simplified, for example the ratio in the r.h.s. of NS equation is
\be  {\eta \over p}={\eta\over s} {4\over T}={1\over \pi T}=1\ee
Of course, our problem has with two temperatures and pressures, $p_i$ and $p_f$ at both $x\rightarrow \pm \infty$:
but as the final parameters do not appear explicitly in the equations in the form used, we use
the initial temperature to set the units.

\begin{figure}[!t]
\begin{center}
\includegraphics[width=7.5 cm]{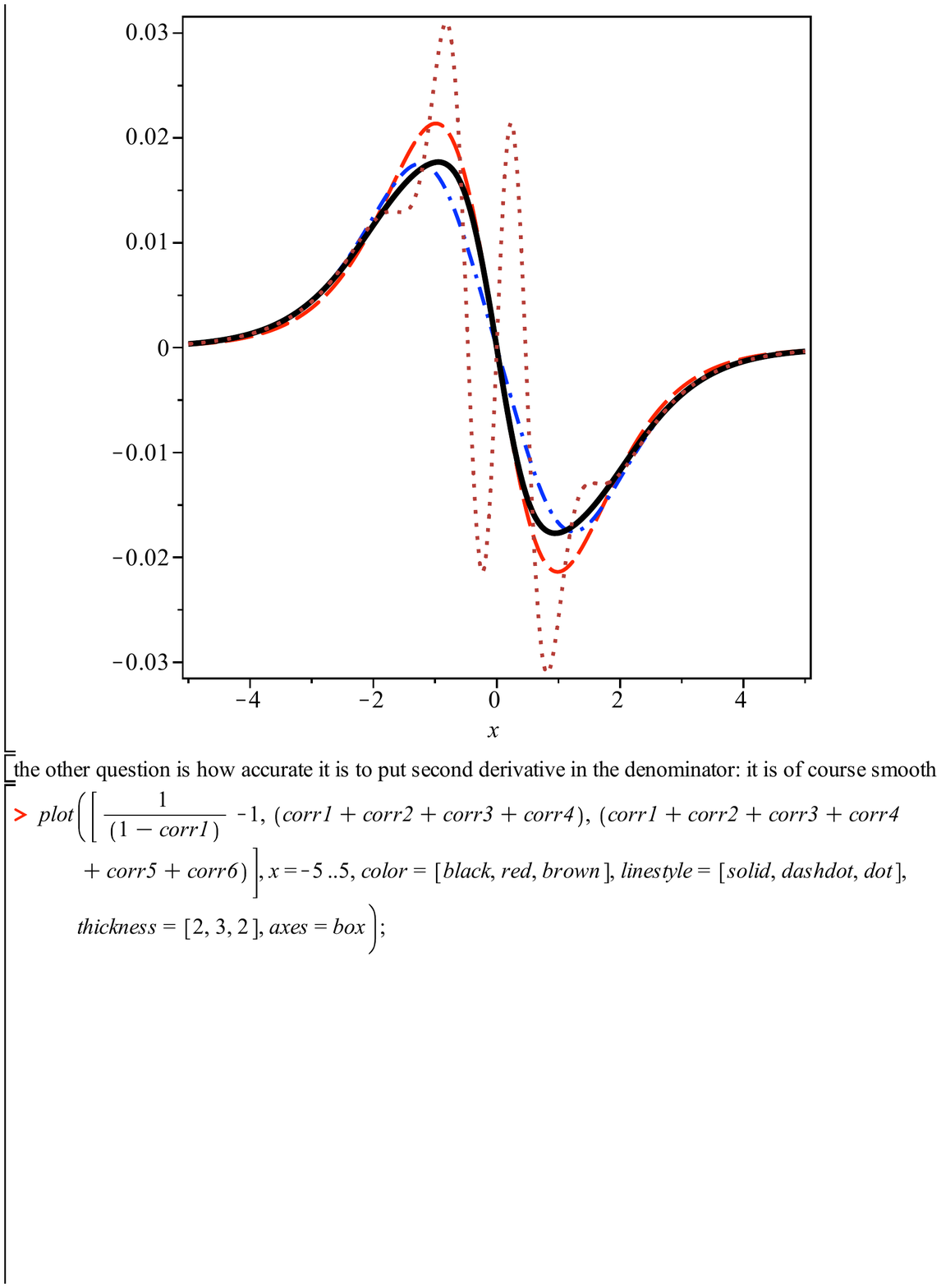}
\includegraphics[width=7.5 cm]{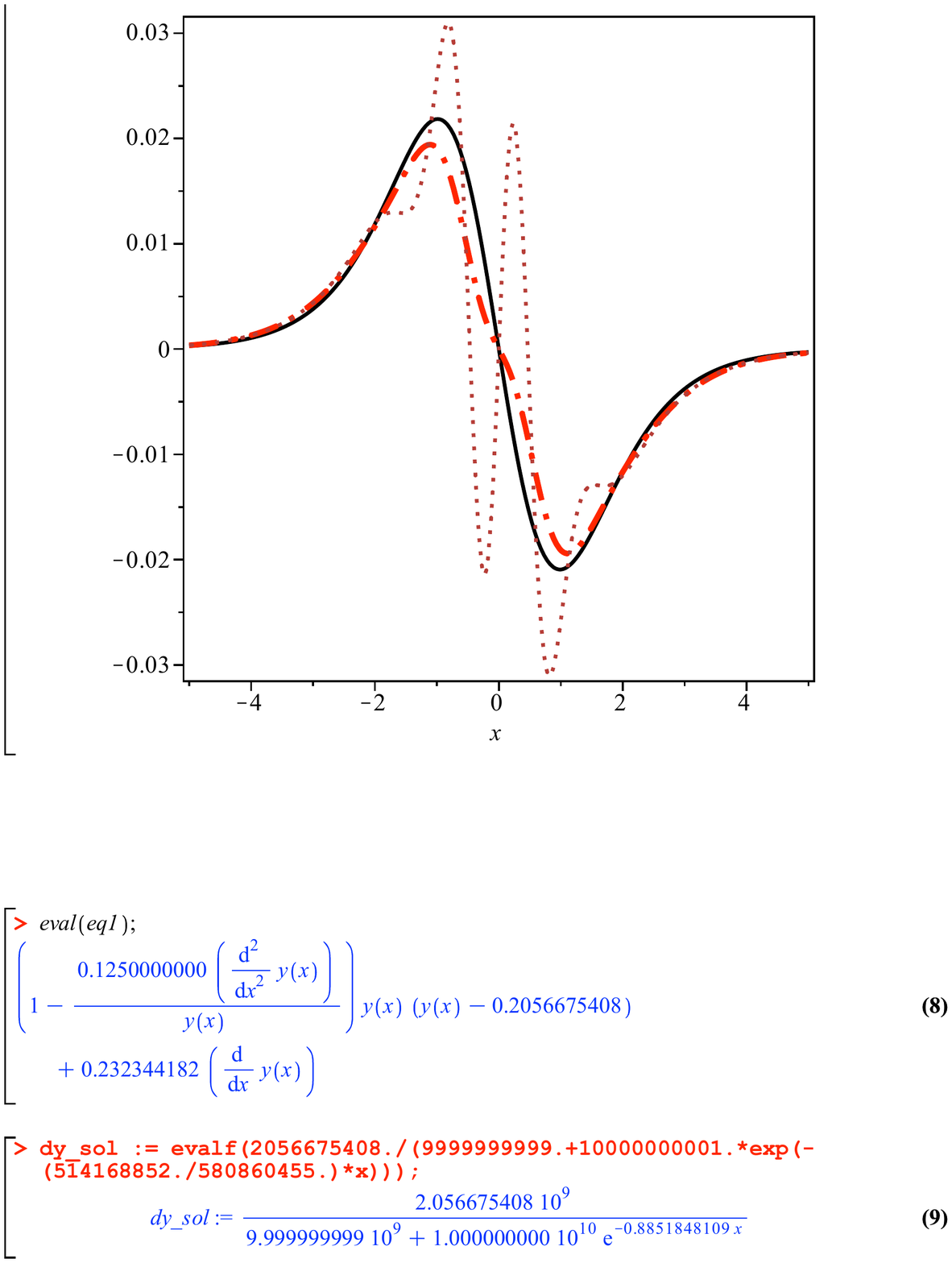}
\end{center}
\vspace{-5ex}\caption{(color online)
(a)  Gradient expansion for the Fermi step with the width $w=0.75$. 
By the (red) dashed, (blue) dash-dotted, (black) solid and (brown) dotted lines we show 
the results including up to 2,4,6 and 12  gradients. 
The lower plot shows the LS modification factor $1/(1-y"/8y)$ applied to the same
shape, by black solid line.  (Red) dash-dotted and (brown) dotted line have up to 8 and 12 derivatives, respectively.
}
\label{fig_LS}
\end{figure}

The first issue to be discussed is the magnitude of the higher gradient terms and convergence
of the derivative expansion. Using for simplicity our LS2 model we write
\be  {1 \over  1- \partial^2/8}=1+   \partial^2/8 + (\partial^2/8)^2+...\ee 
and apply it to the shock profile, taken for illustration to be  the Fermi-like shape
\be 
f={1 \over 1+exp(-x/w)}
\ee
Weak shocks have large width $w>>1$ and thus small gradients, so that the convergence is
parametrically justified. Proceeding to strong shocks, such as our  example displayed in Fig.\ref{fig_NS},
one finds the width  of the order one: thus no small parameter is available.
Few terms in the expansion for this example are shown in Fig.\ref{fig_LS}. The first  thing to notice
is the fact that all corrections are below 1 percent, more than order of magnitude
below the level which  the numerical smallness $1/8$ of the expansion suggest. 

%
  To investigate the issue further, one may reduce the width and check how this expansion in gradients
  behaves. We observed the first signs of trouble at the width $w=0.75$. As seen from the upper Fig.\ref{fig_LS},
 the first terms indicate a reasonably small correction, with the magnitude of 2\%: but going further
 (see the brown curve with 12 derivatives) one finds widely oscillating corrections of increasing amplitude.
 This is a behavior typical for asymptotic series, which converge only till the certain term. It is usually
 assumed in such cases that the last ``good" term shows the best possible approximation: we take the black
 curve including up to 4th correction (8-th derivatives) to by the closest to the truth.
 
 \begin{figure}[!t]
\begin{center}
\includegraphics[width=7.5 cm]{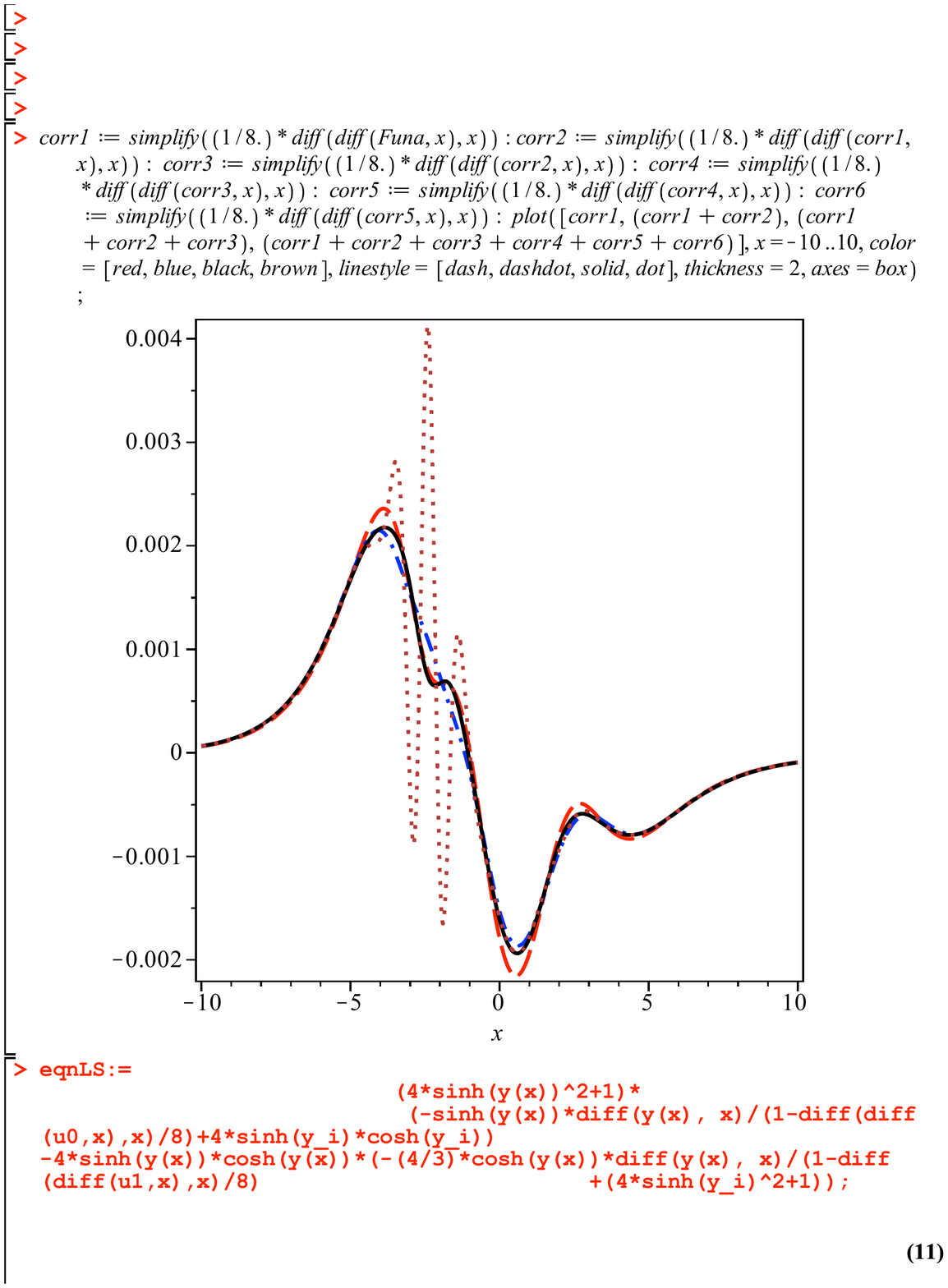}
\includegraphics[width=7.5 cm]{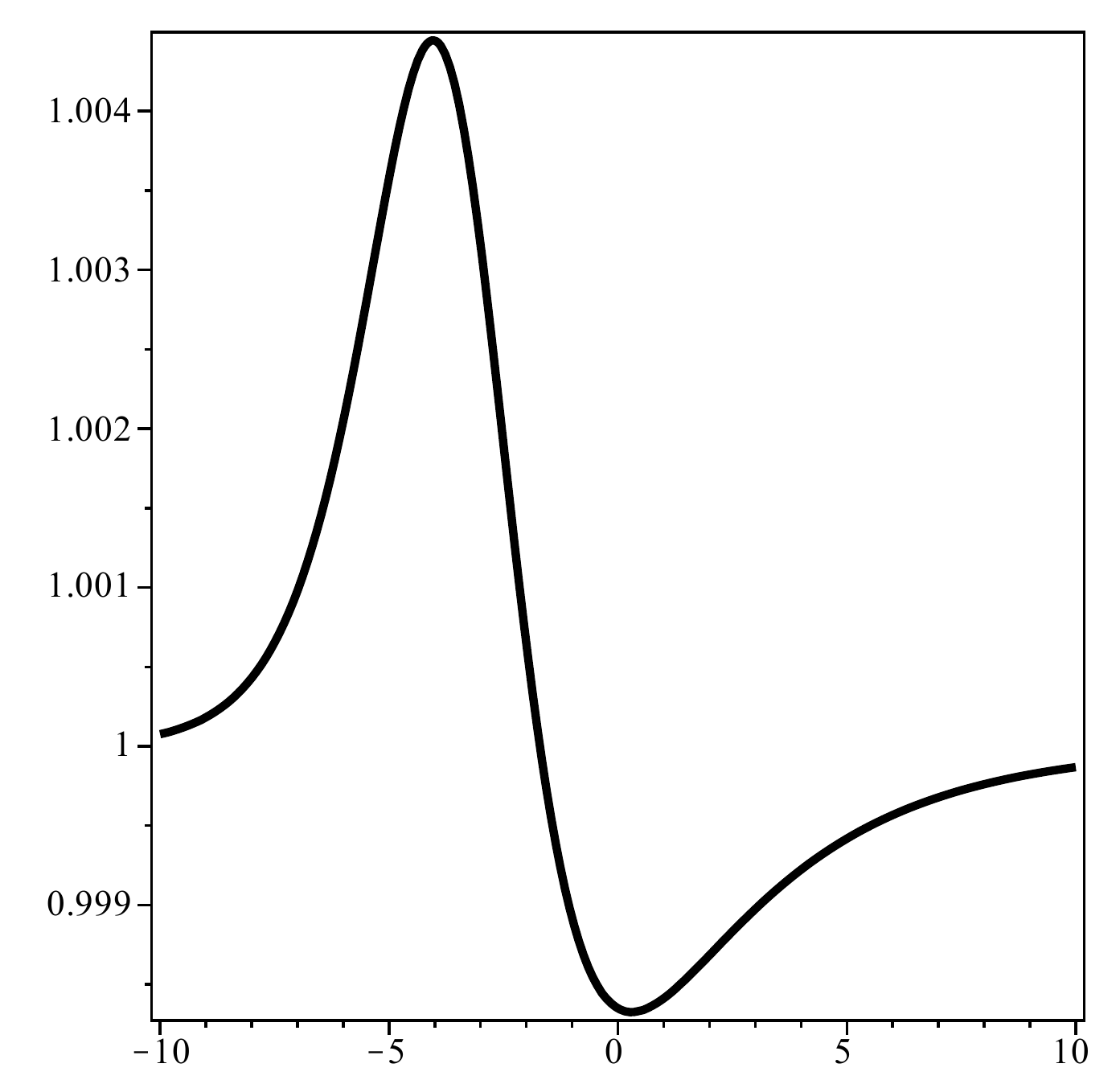}
\end{center}
\vspace{-5ex}\caption{(color online)
The upper plot shows several terms of the gradient expansion for the NS solution shown in Fig.\ref{fig_NS}.
By the (red) dashed, (blue) dash-dotted, (black) solid and (brown) dotted lines we show 
the results including up to 2,4,6 and 12  gradients. 
The lower plot shows the LS modification factor $1/(1-y"/8y)$ applied to the same numerical solution
of the NS equation: its magnitude and shape should be compared to the solid curve in the upper figure.  
}
\label{fig_LS_on_NS}
\end{figure}

 The next issue is if the ``resumed" version in which the first correction (two derivatives)
  is put into the denominator. In other words, the question is if one can naively
 sum the geometric series, ignoring the operator nature of the derivative.  (Of course,  the exact
resumption into the  denominator is provided by the inverse of the differential operator in question.)
 The result is shown in the lower Fig.\ref{fig_LS} by the black curve, which is indeed quite close in shape
 to the sum of the first 4 corrections (the red curve). 

We have also applied the same operations as for the Fermi step functions above to the 
numerical solution of the NS equation (\ref{NS_nonexpanded}) shown in Fig.\ref{fig_NS}.
The results are very much the same as above for the Fermi step function. The upper figure
shows what happens with a series in subsequent gradient expansion: it seems well convergent
first, but the 6-th order term
(12 derivatives) gets oscillating. The lower curve shows the LS resumed  factor, with the second
derivative in the denominator. Note that the whole scale of effect is very small, about 0.003:
we thus conclude that for practical purposes our strong shock example 
shows only negligible effect of the higher gradients. The NS approximation
seem to be justified for this example, which was one of our main results. 

%
%

The conclusion from that is:  the LS resumed hydrodynamics predicts,  that
the sum of the high gradient terms is much smaller than individual ones and basically vanish.
The shock is thus one more example of unexpectedly early onset of the NS behavior.

 It is still somewhat surprising  how small is the LS resumption correction, even for a  relatively strong shock
example under consideration. Part of the explanation is perhaps the observation
that the second derivative vanishes at the inflection point, which is located right at
the center of the shock solution. Such explanation would not be however applicable to
two more  recent examples of rapid onset of the NS regime discussed in Refs.\cite{Chesler:2010bi,janik2}.

\section{Shocks in the  AdS/CFT } 
We use 
coordinates $v, x_1=x, x_2, x_3, r$ and write the nonzero component of the metric as 
\be
g_{11}&=&-r^2 f c^2 +r^2 s^2;  \nonumber \\
g_{12}&=&g_{21}=-r^2 f c s+r^2 c s +A(x_1,r);  \nonumber \\
  g_{22}&=&-r^2 f s^2+r^2 c^2+B(x_1,r) ;          \nonumber \\
                                                   g_{15}&=&g_{51}=c; \nonumber \\
                                                      g_{25}&=&g_{52}=s;             \nonumber \\                                                                                                        g_{44}&=& g_{33}=r^2
\ee
where
\be f&=&1-h(x,r)^4/r^4; \nonumber \\ 
c& =& cosh [y(x)]; \, \, s=sinh [y(x)] \ee
Note that we do look for a static ($v$-independent) solution, depending on only 2 spatial coordinates $r,x$.
The metric contains 4 unknown functions, $y,h,A,B$.

If the two functions $h$ and $y$ are just constants, the Einstein equations (with appropriate cosmological constant)  are satisfied without extra correction, namely with $A,B=0$. Indeed, in this case
 this metric is nothing else but the black  brane moving with the rapidity $y$.
The shock is an $interpolation$ between two such solutions, with $y_i,h_i$ different
from $y_f,h_f$ in a finite-width region, the shock. Needless to say, the values are not arbitrary
and must  be related by
the  continuity of the energy and momentum flux, as we discussed at the beginning of the paper.

The corresponding Einstein equations can be easily derived (e.g. by the ``tensor package" of Maple),
but they look very long and discouraging. Here is  the simplest of them, the Ricci scalar 

\be 
R&=& - (-96 A^2 h^3 c^2 h'' r^2-96 s A h^3 c^3 h'' r^2 B-40 r^8 \nonumber \\
&&+16 s c A^2 r^4 A''-12 r^6 B+4 r^4 A^2-4 A A'' r^6 \nonumber \\
&&+16 c^6 h^4 B A'' A+32 c^2 B r^3 h^3 (h')  \nonumber \\
&&+16 c^4 B^2 h^3 (h') r
-12 c B s r^5 A' \nonumber \\
&&-12 B' r^7+24 c^4 B^2 h^3 h'' r^2 \nonumber \\
&&+2 c^6 h^4 B^2 B''-8 A A'' r^4 c^2 B+4 c r^6 \dot{A}' \nonumber \\
&&+16 r^2 s A^2 \ddot{y} c^3-4 r^2 c^5 B^2 s \ddot{y} \nonumber \\
&&+48 s A^2 c A' r^3+52 c A B' r^5 s+8 s c r^3 h^4 A' \nonumber \\
&&+8 c^2 r^3 s A \dot{A}+16 c^2 r h^4 A A'-4 c^4 r B h^4 B' \nonumber \\
&&-16 c^4 r A A' h^4-52 A A' r^3 c^2 B-64 A^2 h^3 c^2 (h') r \nonumber \\
&&+64 A^2 h^3 c^4 (h') r+72 s A c B r^4-8 s A h^4 r^2 c \nonumber \\
&&-16 s A c^3 B h^4-2 r^8 B''+3 r^6 \dot{y}^2-3 r^6 A'^2 \nonumber \\
&&+16 A^4 c^2-16 A^4 c^4-16 s c^5 h^4 A^2 A'' \nonumber \\
&&-4 s A r^5 \dot{y}-6 c^2 B A'^2 r^4-10 c^4 B \dot{y}^2 r^4 \nonumber \\
&&+20 c^3 B r^5 \dot{y}+2 c^2 B \dot{y}^2 r^4+16 A^3 c^3 s B \nonumber \\
&&+24 c^6 B^3 h^2 (h')^2-4 c^6 h^4 B A'^2-c^6 h^4 B B'^2 \nonumber \\
&&+4 c^4 h^4 B A'^2+2 s r^6 \dot{y} A'-r^2 c^4 h^4 B'^2 \nonumber \\
&&+4 r^2 c^2 h^4 A'^2-16 r A^2 \dot{A} c^3+16 r^3 c^5 B \dot{A} \nonumber \\
&&+16 A^2 r^3 c^5 \dot{y}+r^4 c^4 B B'^2-4 c^4 h^4 r^2 A'^2 \nonumber \\
 &&+4 A^2 r^2 c^2 A'^2+16 A^2 r^2 c^4 \dot{y}^2 \nonumber \\
 &&-4 A^2 r^2 c^2 \dot{y}^2-48 r c^4 A^3 A'+48 r c^2 A^3 A'- \nonumber \\
 &&4 r^2 c^4 A^2 A'^2-12 r^2 c^6 A^2 \dot{y}^2+16 c^5 r A^2 \dot{A}\nonumber \\
 &&+8 c^5 r B^2 \dot{A}+4 r^3 c^5 B^2 \dot{y}-8 r^3 c^4 B^2 B' \nonumber \\
 &&-3 r^2 c^6 B^2 \dot{y}^2-3 r^2 c^4 B^2 A'^2 \nonumber \\
 &&-96 r^2 c^4 B A^2+8 r^5 c^4 B B'-r^2 c^4 B^2 \dot{y}^2 \nonumber \\
 &&+16 r^3 c^5 A \dot{B}-56 A^2 r^3 B' c^4-4 c^4 r^4 \dot{y} \dot{A} \nonumber \\
 &&+4 c^2 r^4 \dot{y} \dot{A}-2 c^3 r^4 A' \dot{B} \nonumber \\
 &&+12 c^3 B^2 \dot{y} r^3-32 c^2 r^5 A' A-8 A^2 \dot{y} r^3 c^3  \nonumber \\
 &&-2 r^6 c^3 \dot{y} B'+2 r^6 c \dot{y} B'+32 r^5 c^4 A A' \nonumber \\
 &&-32 s c^5 A B h^3 (h') B'-32 r^2 s A c^3 h^3 (h') B' \nonumber \\
 &&-32 r^2 c^3 B s A' h^3 (h')-2 r^2 c^4 B A s \dot{y} B' \nonumber \\
 &&-288 s A h^2 c^3 (h')^2 r^2 B+16 c^4 r^2 B h^3 (h') B' \nonumber \\
 &&+64 c^4 r^2 A A' h^3 (h')-64 c^2 r^2 A A' h^3 (h') \nonumber \\
 &&+4 c^2 B s \dot{y} r^4 A'-144 s A h^2 c (h')^2 r^4 \nonumber \\
 &&+64 s A^2 c^3 A' h^3 (h')-64 c^4 B A A' h^3 (h') \nonumber \\
 &&-64 s c^5 A^2 A' h^3 (h')+4 s c^5 h^4 B A' B' \nonumber \\
 &&144 s A c^5 B^2 h^2 (h')^2+64 c^6 B A A' h^3 (h') \nonumber \\
 &&-16 c^5 B^2 h^3 s (h') A'+4 r^4 c^4 B s \dot{y} A' \nonumber \\
 &&-2 r^4 c^3 B s A' B'+8 r^2 c^3 B A A' \dot{y}-24 c^4 r s A \dot{A} B \nonumber \\
 &&+2 r^2 c^4 B A A' B'+2 r^2 c^4 B^2 s \dot{y} A'  \nonumber \\
 &&+4 r^2 c^4 B s A' \dot{A}+2 r^2 c^5 B s \dot{y} \dot{B} \nonumber \\
 &&-4 r^2 c^5 B A A' \dot{y}+4 r^4 s A c^4 \dot{y} B' \nonumber \\
 &&-6 r^4 s A c^2 \dot{y} B'-8 r^2 s A c^3 \dot{y} \dot{A} \nonumber \\
 &&-8 r^2 s A^2 c^2 A' \dot{y}-16 s r^4 h^3 c (h') A' \nonumber \\
 &&+12 s c^5 r^2 A \dot{y}^2 B+44 s c^3 r^3 A B B' \nonumber \\
 &&+4 s c^4 r^2 A A' \dot{B}+8 s c^3 r^2 A B A'^2 \nonumber \\
 &&+8 s c^5 r^2 A \dot{y} \dot{A}-4 s c^3 A^2 r^2 A' B' \nonumber \\
 &&+56 s c^3 A^2 r A' B+4 s r^2 h^4 c^3 A' B'-16 s c^4 r^3 B A \dot{y} \nonumber \\
\ee
\be
 &&-28 c^2 B s A \dot{y} r^3-4 r^6 c s \ddot{y}+2 r^4 c^2 h^4 B'' \nonumber \\
 &&-16 r^2 A^2 \dot{A}' c^3-2 r^4 c^4 B^2 B''-16 r^2 c^4 A^3 A'' \nonumber \\
 &&-96 A^2 h^3 c^4 h'' B+8 r^6 h^3 h''-4 r^6 B'' c^2 B \nonumber \\
 &&-48 s A h^3 c h'' r^4-48 s A c^5 B^2 h^3 h''-8 c^5 h^4 B B'' s A \nonumber \\
 &&+8 c^6 h^4 A^2 B''+8 r^4 c^3 B B'' s A-8 s c^3 h^4 B A'' r^2 \nonumber \\
 &&-16 r^2 c^4 B \ddot{y} A-16 c^2 r^2 h^4 A'' A+96 A^2 h^3 c^4 h'' r^2 \nonumber \\
 &&+16 c^4 r^4 \ddot{y} A-16 c^2 r^4 \ddot{y} A+8 c^2 A^2 B'' r^4 \nonumber \\
 &&+16 c^4 r^2 h^4 A'' A+224 c^2 A^2 r^4-8 c^4 h^4 A^2 B'' \nonumber \\
 &&+8 c^6 B^3 h^3 h''-16 s r^5 c^3 B A'-16 s r^5 c^3 A B' \nonumber \\
 &&+4 s r^6 \dot{y} A' c^2+8 r^2 A c^3 A' \dot{A}+4 r^2 A c^4 \dot{y} \dot{B}  \nonumber \\
 &&-8 r^4 c^5 A A' \dot{y}+4 r^4 c^2 s A' \dot{A}+2 r^4 c^3 s \dot{y} \dot{B}  \nonumber \\
 &&-8 r^2 c^5 A A' \dot{A}-4 r^2 c^6 A \dot{y} \dot{B}+4 r^4 c^4 A A' B' \nonumber \\
 &&-2 r^4 c^3 A s B'^2-8 r c^3 A^3 B' s+4 r c^4 A^2 B' B  \nonumber \\
 &&+4 r^2 c^5 A^2 \dot{y} B'+96 r^2 s A c^3 B^2-8 r^3 c^4 B s \dot{B}  \nonumber \\
 &&-4 r^2 c^6 B \dot{y} \dot{A}-2 r^2 c^5 B A' \dot{B}+24 r^3 c^4 B A A' \nonumber \\
 &&-16 r c^4 B^2 A A'+8 c^4 r A^2 \dot{B} s-4 c^5 r A \dot{B} B \nonumber \\
 &&-12 s c^3 r^3 B^2 A'+8 r^4 A A' c \dot{y}-2 s r^6 c A' B' \nonumber \\
 &&+4 r^2 c^4 B \dot{y} \dot{A}-32 r^3 s A \dot{A} c^4-16 A^3 r c^2 \dot{y} s \nonumber \\
 &&+8 A^2 r c^3 \dot{y} B+224 r^4 c^3 B A s-2 r^4 c^5 B \dot{y} B' \nonumber \\
 &&-4 A^2 r^2 c^3 \dot{y} B'+8 r^4 c^2 h^3 (h') B'+4 c^3 r^4 A A' \dot{y} \nonumber \\
 &&-2 A c^2 A' r^4 B'-288 A^2 h^2 c^2 (h')^2 r^2 \nonumber \\
 &&+288 A^2 h^2 c^4 (h')^2 r^2+2 c^3 B \dot{y} r^4 B'-8 c^3 s h^4 A A'^2 \nonumber \\
 &&+192 A^3 h^2 c^3 (h')^2 s-288 A^2 h^2 c^4 (h')^2 B+32 c^6 A^2 h^3 (h') B' \nonumber \\
 &&+8 c^5 s h^4 A A'^2+2 c^5 s h^4 A B'^2+288 c^6 B A^2 h^2 (h')^2 \nonumber \\
 &&+8 c^6 B^2 h^3 (h') B'-8 c^6 h^4 A A' B'-192 s c^5 A^3 h^2 (h')^2 \nonumber \\
 &&+8 c^4 h^4 A A' B'-32 A^2 h^3 c^4 (h') B'+8 s A c A'^2 r^4 \nonumber \\
 &&+20 s A c^3 \dot{y}^2 r^4-40 s A c^2 r^5 \dot{y}-8 s A c \dot{y}^2 r^4 \nonumber \\
 &&+72 c^4 B^2 h^2 (h')^2 r^2+72 c^2 B h^2 (h')^2 r^4 \nonumber \\
 &&-64 s A c^3 B h^3 (h') r+8 s r^6 B'' c A+16 r^2 c^6 B \ddot{y} A \nonumber \\
 &&+192 s A c r^6+12 c B r^5 \dot{y}+4 c^2 B h^4 r^2 \nonumber \\
 &&-16 s c r^7 A'-20 c^3 r^3 A \dot{B}-8 c^2 r^5 s \dot{B} \nonumber \\
 &&-20 c^2 r^5 B B'+60 c^2 A^2 B' r^3-4 r^3 c^2 h^4 B' \nonumber \\
 &&+96 A^2 r^2 c^2 B-64 s A r^3 h^3 (h') c+8 s c^3 h^4 B A' r \nonumber \\
 &&+8 c^3 s h^4 A B' r+64 A^3 h^3 c^3 h'' s-4 s c^5 h^4 B^2 A'' \nonumber \\
 &&-16 c^4 h^4 B A'' A-7 c^2 r^6 \dot{y}^2+r^6 c^2 B'^2 \nonumber \\
 &&+24 r^6 h^2 (h')^2-8 c r^5 \dot{A}-4 A^2 c^4 B^2 \nonumber \\
 &&-56 r^4 c^4 B^2-24 r^2 c^4 B^3-224 A^2 r^4 c^4 \nonumber \\
 &&+16 r^2 c^5 A^2 \dot{A}'-8 r^4 c^4 A^2 B''+16 r^2 c^2 A^3 A'' \nonumber \\
 &&+4 r^2 c^5 B^2 \dot{A}'+8 c^3 r^4 \dot{A}' B+4 c^4 h^4 B r^2 B'' \nonumber \\
 &&+16 r^5 h^3 (h')+16 r^7 c \dot{y}+16 c^3 r^5 \dot{A} \nonumber \\
 &&+8 c^2 r^7 B'+16 A^2 c^4 h^4-16 A^2 c^2 h^4 \nonumber \\
 &&-12 A A' r^5-96 c^2 B r^6-36 c^2 B^2 r^4+4 c^4 B^2 h^4 \nonumber \\
 &&+24 c^2 B h^3 h'' r^4-64 s c^5 A^3 h^3 h''-16 r^4 c^2 s \dot{A}' A \nonumber \\
 &&-16 s c^4 r^2 A \dot{A}' B+96 c^6 B A^2 h^3 h''-16 s c^5 r^2 A^2 \ddot{y} \nonumber \\
 &&-4 s c r^4 h^4 A''+16 s c^3 A^2 r^2 A'' B-8 c^3 s h^4 A r^2 B'' \nonumber \\
 &&-8 r^4 c^3 s \ddot{y} B+16 s c^3 h^4 A^2 A''-4 r^2 c^4 B^2 A A'') \nonumber \\
 && {1 \over 2r^2 (r^2-2 c s A+c^2 B)^3} 
\ee
where the prime stands for the derivative over $r$ and the dot means the derivative over $x_1$. 
Substituting again the moving brane solution one gets constant (coordinate-independent) value $R= 20$, related
to the AdS cosmological constant.

\begin{figure}[!t]
\begin{center}
\includegraphics[width=7 cm]{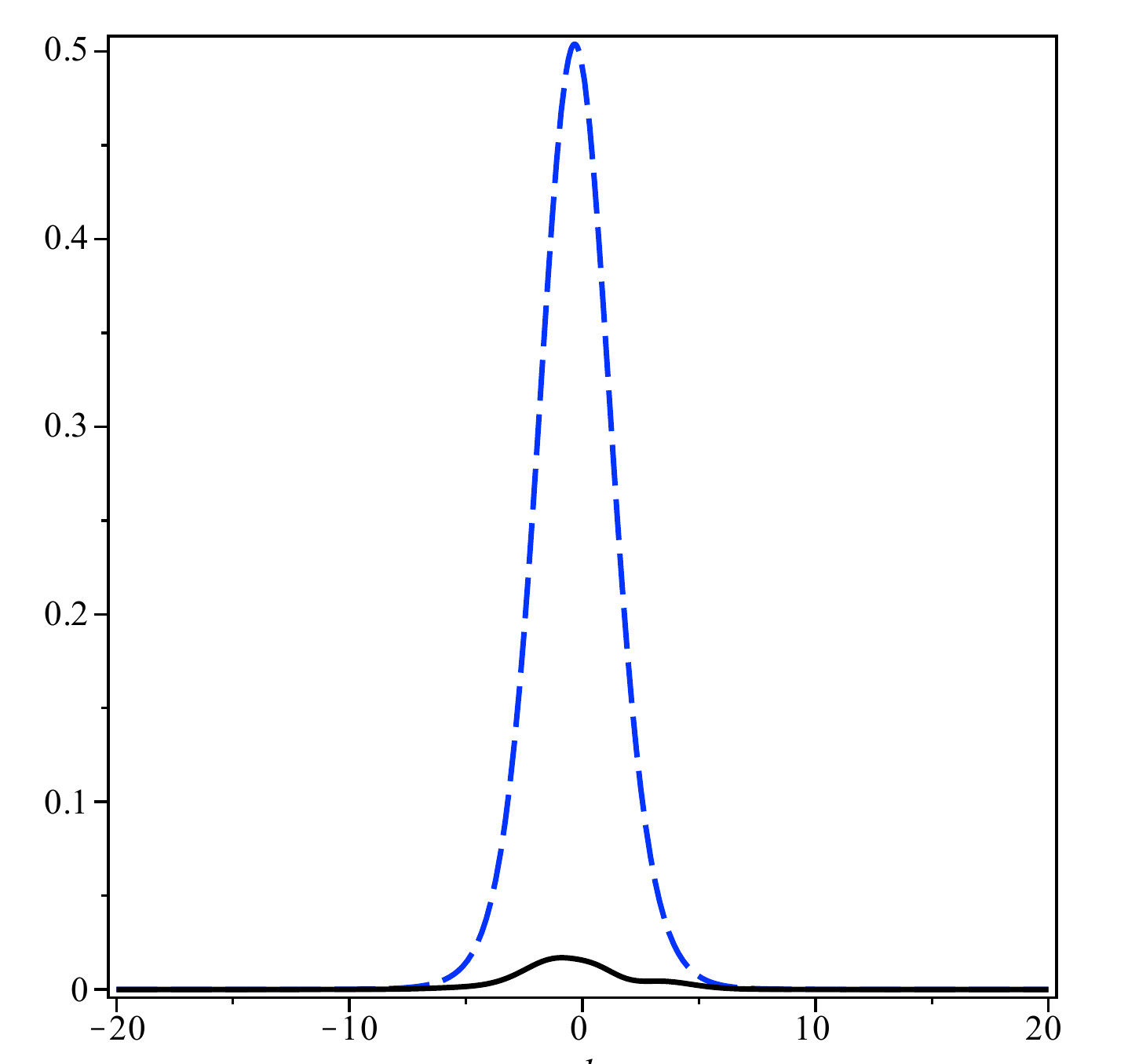}
\includegraphics[width=7 cm]{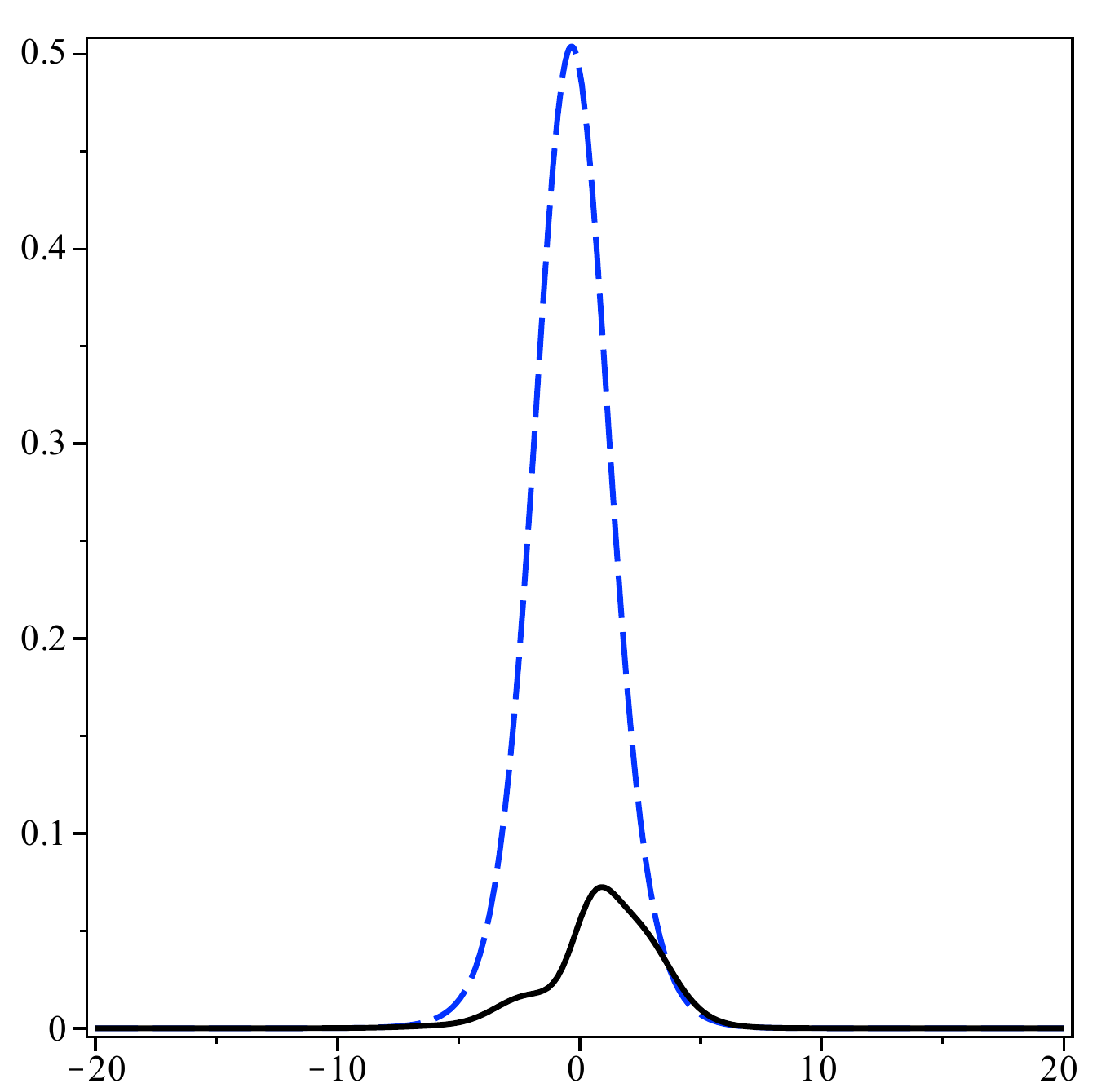}
\end{center}
\vspace{-5ex}\caption{(color online) 
The scalar square of the modified Einstein tensor (\ref{eqn_ES}) (upper plot)
and the
simple sum of squares of all components (lower plot) the as a function of the coordinate $x$
(in units of $1/\pi T_i$), for $r=1$.  The dashed (blue) curve is for 
the case without the $B$ correction (\ref{eqn_corr}), the solid (black) line  includes it .  }
\label{fig_sum}
\end{figure}

   One can start by taking drastic simplifications: putting extra functions to zero $A,B=0$ and $h(x_1,r)=h(x_1)$
to be independent on $r$. This results in the following relatively simple expression
\be R-20&=& {1\over 2 r^2}[-16 r cosh(y) \dot{y}+ \dot{y}^2(7 cosh(y)^2  -3) \nonumber \\
&&+4 cosh(y) sinh(y) \ddot{y} ]
\ee
depending only on the rapidity profile across the shock. Let us see what this combination is for
traditional Fermi-step solution in weak shock Example 1. Since in this case the width of the shock is large,
the square of the derivative and the second derivative are negligibly small, while only the first term is
important. With this lesson in mind, one can return to a case with nonzero extra functions  and use linearized  equations,
 ignoring higher derivatives in $x$. This procedure leads to hydrodynamics, 
as one can  solve all the equations for extra function perturbatively, see \cite{Hubeny:2011hd} for general
review of the method and Refs \cite{1004.3803,1105.1355} for the particular case of weak shocks.

Let us turn to strong shocks, such as our numerical example, in which case the width of the shock is $O(1)$ and
there is no small parameter in the problem.  The Einstein equations in full nonlinear form take pages and are way too complicated
to be presented here or solved directly. Instead we propose to use a variational method.

Unfortunately the Einstein-Hilbert action $R$ is not bounded from below and cannot be used for variational
studies. The so called conformal gravity, with a squared Weyl tensor in the Lagrangian, should work \cite{comment}.
 What we propose to do is to use the covariantly squared (modified) Einstein tensor 
\be 
\bar{E}^2=\bar{E}_{mn} \bar{E}^{mn},\,\,\, \bar{E}_{mn}=E_{mn}+6 g_{mn}
\label{eqn_barE}
\ee
which combines all the Einstein equations (in the AdS/CFT setting) into one (covariant
scalar) combination.  For a check, one can of course look at all individual components of the modified Einstein tensor. We used for this purpose a (non-covariant!) sum of squares
\be ES=\sum_{mn} (\bar{E}_{mn} )^2  \label{eqn_ES} \ee
This combination is of course sign-definite, and since for a solution all components, and thus the sum of squares, should vanish, it can be used to monitor the ``variational progress" for all of the components.

The equations are basically elliptic, and as such they only need the field values on the boundary
of the region to be appropriately interpolated (solved) inside it. In this problem all corrections vanish both at $x\rightarrow \pm \infty$,
as well as at large $r$. The only tricky issue is near the nontrivial boundary  of  the black hole horizon. We use
as the initial input the rapidity $y(x)$ and pressure $p(x)=h^4(x)$ from the numerical NS solution, 
which fix the $g_{vv}$, but allow nonzero
modification functions in other components such as $g_{xx}$. The inter mediate step we will not dwell on
was a parameterization of the solution in simple enough form, so that the evaluation of the huge
expressions involved be possible.

A simple way to proceed is to use
 variationally,  to use certain ansatz (assumed trial function) and
substitute it into all the equations and/or $\bar{E}^2$ see how close/far are the results from the desired zero values.
Evaluation of a not-too-complex trial function is performed by Maple in  seconds,
 in spite of horrendously complex
expressions involved. 
In Fig.\ref{fig_sum}(a)  one can see a comparison of the $\bar{E}^2(x,r)$ for the NS profile only
(the top blue curves) with the results including this $B$ (the lower black curves).
Note first, that even the dashed curve which use only the input rapidity and pressure
hydrodynamical solution (shown in Fig.\ref{fig_NS}) is already not too bad. Indeed,
the  components of the Einstein tensor are expected to be $O(1)$, as the problem 
has no parameters, and the number of nonzero ones summed up is 11 (9 from the $t,x^1,x^5$ block plus
$\bar{E}_{33}=\bar{E}_{44}$, other 6 are zero due to parity of $x^3$ and $x^4$).  Thus one expects
something of the order $O(10)$ and gets something 20 times less at its peak. 

We have used only one of the correction functions $B$, and after some number of trials we 
came up with 
 the flowing ansatz for it
\be 
B(x,r)=-0.052 r (1-0.3 x) exp[-.1( x+0.3)^2]
\label{eqn_corr}
\ee
It is clear that the mismatch is reduced by another order of magnitude, in the whole region of
$x,r$ in question. (We don't show it as a function of $r$, but the relative error is constant in $r$).
One can  view this  mismatch as the amount of external matter (or, more exactly, stress) which
is needed to make our approximate solution of the Einstein equation exact.

In the next Fig.\ref{fig_sum} (b) we plot the simple sum of squares of all components of the (modified) Einstein
tensor. While we have not minimized this quantity, it also shows  a significant reduction of the mismatch. 
We take it as the indication that there are no significant cancellations between components
in the scalar and 
we are in fact close to the solution, to which one can get even closer with
 more sophisticated functions used.
The standard procedure would be to discretize the functions, by introducing
a grid in $x_1,r$  and use well known relaxation methods (solving for zero) at each point \cite{comment2} :
in this case one can reach arbitrary high precision if needed.

\begin{figure*}[!t]
\begin{center}
\includegraphics[width=6.5cm]{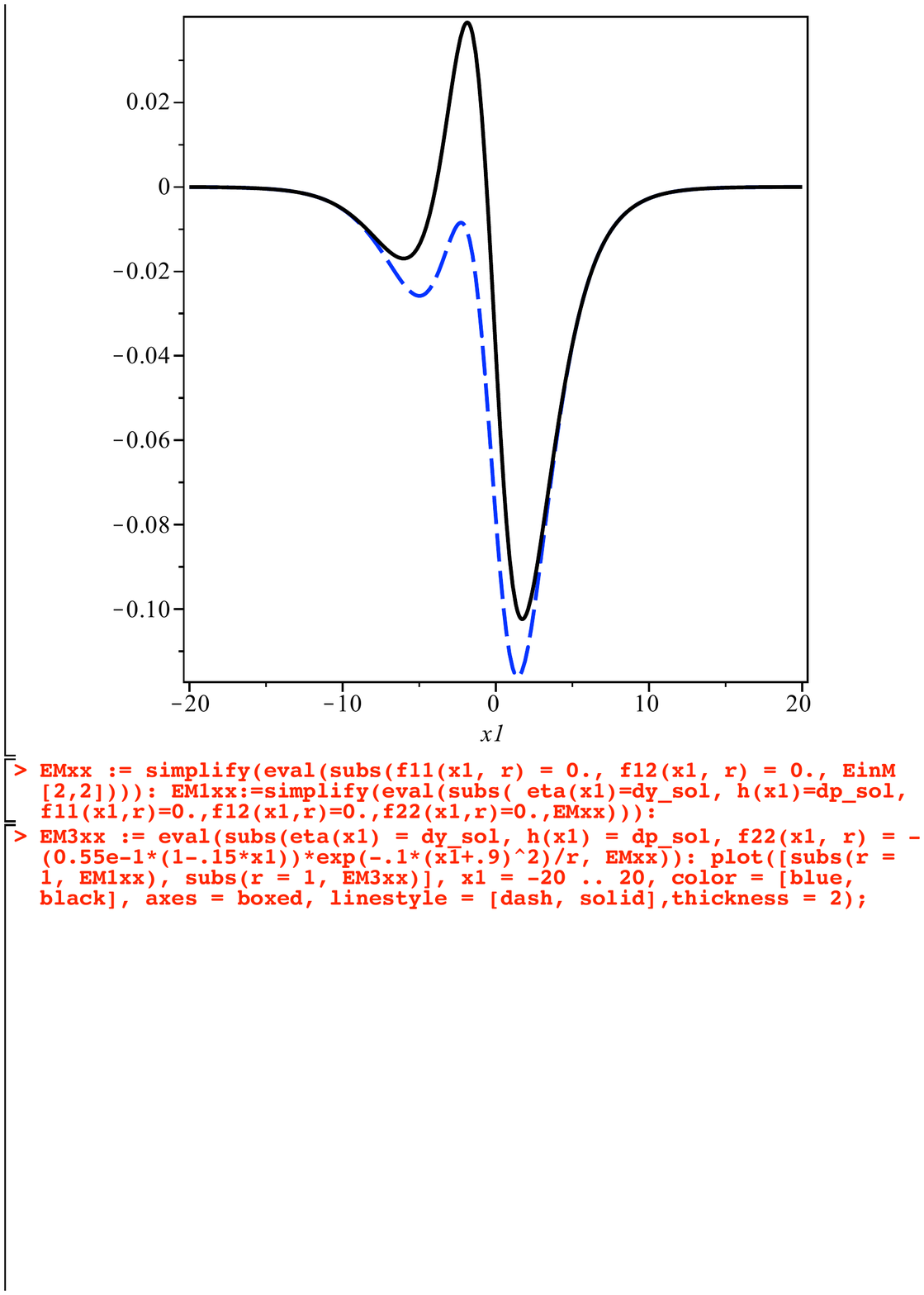}
\includegraphics[width=6.5 cm]{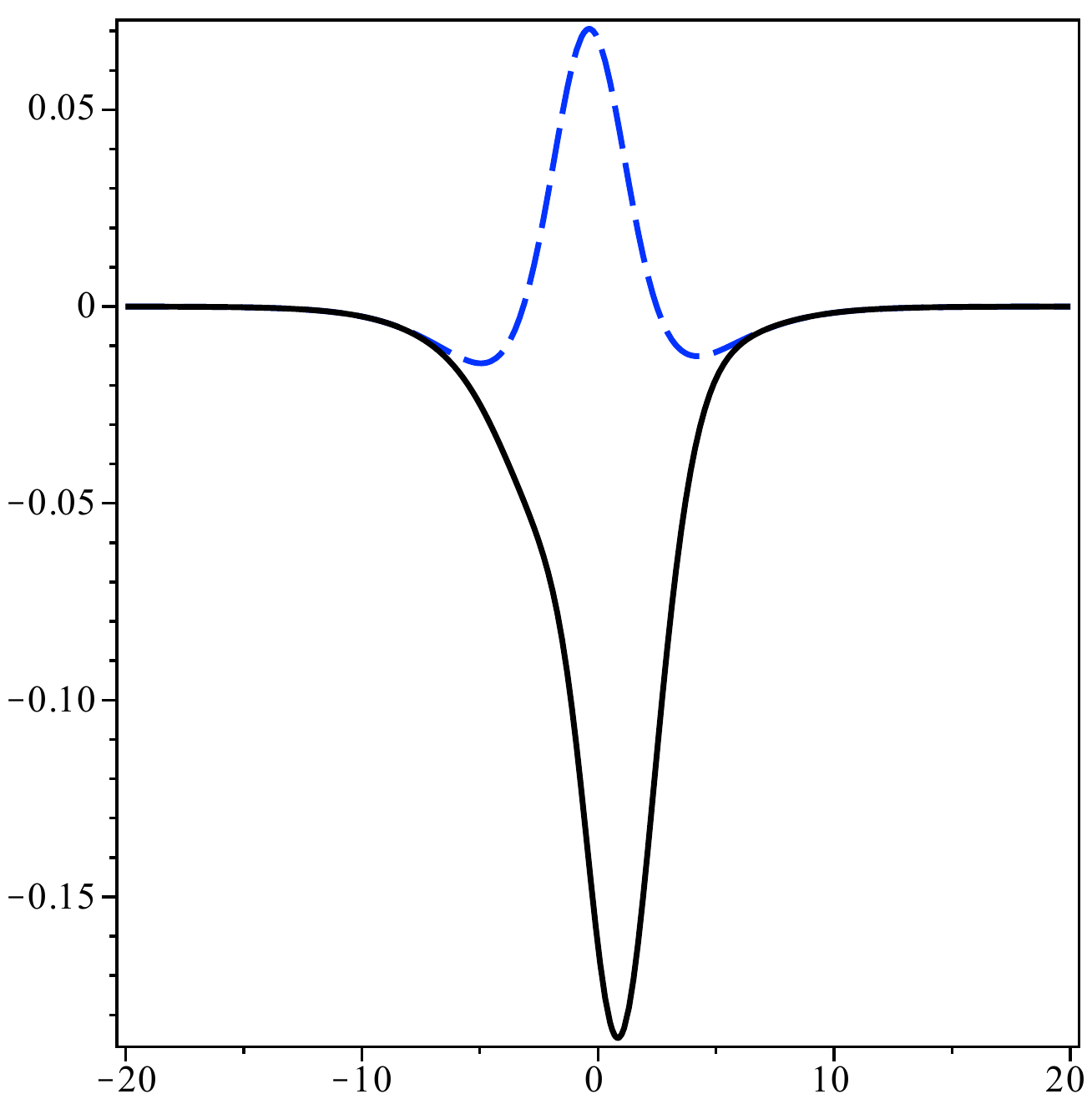}\\
\includegraphics[width=6.5 cm]{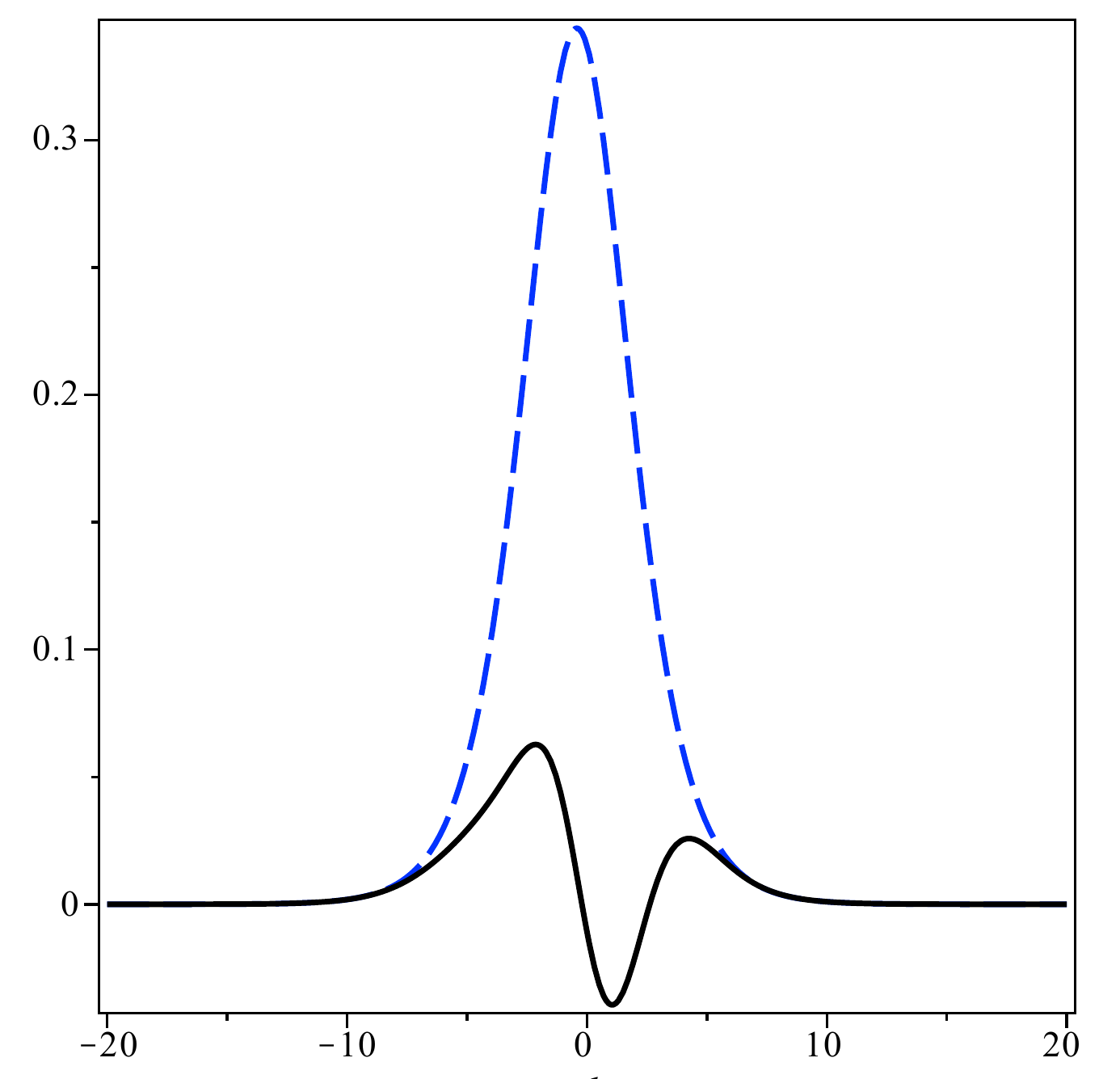}
\includegraphics[width=6.5 cm]{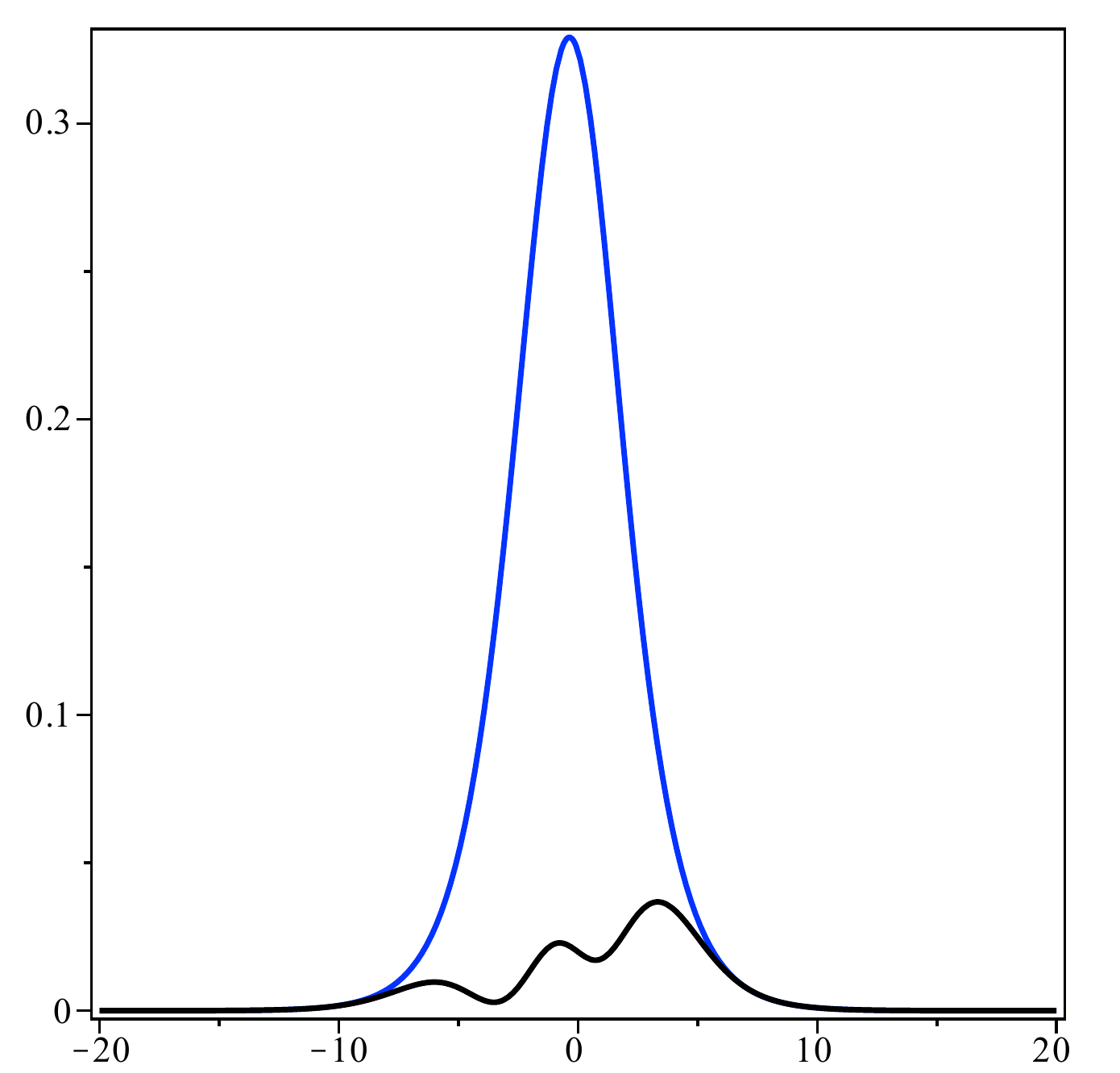}
\end{center}
\vspace{-5ex}\caption{(color online) 
The same notations as for the previous plot, for individual components of the modified Einstein tensor,
from top to bottom $\bar{E}_{11} , \bar{E}_{22}, \bar{E}_{15}, \bar{E}_{33}   $
}
\label{fig_comp}
\end{figure*}

Of course, in the calculation we also were looking at the individual components
of the modified Einstein tensor. Few are shown at Fig. \ref{fig_comp}:
the reader can see that while in some cases ( e.g. the 15,33 components) the improvement is obvious,
other components do not show that. The reader then is invited to look at the scale of all
the graphs and notice that ``bad" ones are much smaller than ``good" ones naturally.
Another way to see that is to note that the difference between the dashed and solid curves
is due to correction $B$ which was found to be only $0.05$ in magnitude, while 
the metric components are generally not small  $O(1)$. 

The meaning of  the correction $B$ is seen from the fact that it modifies
 the length elements along the coordinate $x$ 
\be dl=  \sqrt{g_{xx}} dx  = r \sqrt{-f s^2 +c^2 +B(x,r)/r^2}dx  \ee
Since $B/r^2\sim 1/r$ one finds that far from the black hole the correction is unimportant and thus at the boundary
this correction disappears. Near the horizon, since the first term is small, $f\approx 0$ the second is dominant and $O(1)$.  As the correction $B$ 
 is negative, it shrinks a bit the  distance across the shock. The magnitude of $B$ found
 leads to our main conclusion,  that $B$ makes a shock 
 {\em few percent sharper}  near horizon, as it is at the AdS boundary. 


(Additional comment  about a horizon. In general, as new solution is found one has to calculate the
null geodesics in it and find their bifurcation. Note however that  for ``radial" ones, with  dx=0,  the sign of the $dr/dv$ is defined by the metric components $g_{11},g_{15}$,   
which do not include the modification function $B$. So the line of horizon is still given by the (lorentz transformed)
$h^4(x)$ line.)


\section{ Summary and discussion }

Hydrodynamics  is an approximate effective theory, its local approximation 
relies on smallness of the gradients.  Another way to explain the assumption:  the spatial scale
of interest is much larger than the ``micro scale" defined by the viscosity.  (If kinetic theory
be applicable, the latter would be the mean free path of quasiparticles: but in our problem 
we have neither quasiparticles nor kinetic description.)
Very weak shocks are basically sounds, the gradients are parametrically small and for them  hydrodynamical assumptions are
justified.

 Strong shocks in general have no reasons to be treated in the NS
approximation: there are no small parameter available and  the higher gradient terms are in general not small.
And yet we found,  from two different (although  related) sources, that all 
corrections to NS are of the magnitude of {\em few percents} or less. 

 While for most applications
one usually don't even know the coefficients of the higher gradients, in the 
case of conformal plasmas the AdS/CFT correspondence tells us their values,
and several of them has been calculated.
Lublinsky and myself \cite{Lublinsky:2009kv} (LS)  had proposed to
resum those into some ``universal hydro" form. The alternating signs of the corrections basically lead to
strong cancellation of those higher order terms,  even when they are
not small individually. 
Explicit calculation of the higher derivative terms show that the series are asymptotic. The
LS resumed correction indeed provides results which are very consistent with the 
convergent part of the gradient expansion. They are also small, indicating that solutions to
``universal LS hydrodynamics"  are remarkably close to the NS ones, at least in the case of shocks.

 The second approach used is the AdS/CFT. While looking for  shock-related stationary solutions to the Einstein equations 
we made  the first steps, by using some original form of the variational approach. We have been able to reduce that scalar square of modified Einstein tensor to a fraction of a percent, in the whole range of variables, 
which perhaps indicate we are numerically close to a true solution.  Unlike earlier studies  \cite{1004.3803} , 
there is no expansion in gradients or linearization of the equations in our approach:
 all terms are kept and all of them are individually O(1).
 
Our variational solution finds that one needs  only {\em a few percent correction} to such observables as
the shock width and profile, as compared to the one obtained from the NS equations.
Our results agree in spirit with finding from
two more recent examples  \cite{Chesler:2010bi,janik2} of the onset of the LS/NS hydrodynamics.
 Those two examples are however time-dependent collision problems, which are much more complicated
 technically than the one described in this work.

Needless to say, a lot of work needs to be done. The variational solution should be extended to a better accuracy.
 The  issue may be pursued more into the realm of even stronger shocks. 
  Last but not least,  one now may argue that production of such  shocks  
by strongly quenched jets at RHIC/ LHC heavy ion collisions  opens a possibility to test 
some of these predictions experimentally.
The first steps perhaps would be 
observation of the corresponding Mach cone angles in jet-hadron correlations 
and derivation of the shock wave velocities. 

\vskip .25cm {\bf Acknowledgments.} This work was 
done while I was on sabbatical leave at IAS, Princeton: its support is greatly appreciated.
Discussions with F.Pretorius and J.Maldacena were very helpful.



\begin{thebibliography}{99}
\bibitem{LL} L.D.Landau and E.M.Lifshitz, Fluid mechanics,

\bibitem{LandauBelenky} L.D.Landau and S.Z.Belenky, UFN 56 (1955) 309, also reprinted in Landau's collected works, V2.
\bibitem{Teaney:2000cw} 
  D.~Teaney, J.~Lauret and E.~V.~Shuryak,
  Phys.\ Rev.\ Lett.\  {\bf 86}, 4783 (2001)
  [nucl-th/0011058].


\bibitem{deflagration}
  D.~H.~Rischke and M.~Gyulassy,
  Nucl.\ Phys.\ A {\bf 608}, 479 (1996)
  [nucl-th/9606039].



\bibitem{Teaney:2001av} 
  D.~Teaney, J.~Lauret and E.~V.~Shuryak,
  nucl-th/0110037.

\bibitem{Huovinen:2001cy} 
  P.~Huovinen, P.~F.~Kolb, U.~W.~Heinz, P.~V.~Ruuskanen and S.~A.~Voloshin,
  Phys.\ Lett.\ B {\bf 503}, 58 (2001)
  [hep-ph/0101136].

\bibitem{Hirano:2002ds} 
  T.~Hirano and K.~Tsuda,
  Phys.\ Rev.\ C {\bf 66}, 054905 (2002)
  [nucl-th/0205043].


\bibitem{Teaney:2003kp} 
  D.~Teaney,
  Phys.\ Rev.\ C {\bf 68}, 034913 (2003)
  [nucl-th/0301099].
\bibitem{Romatschke:2007mq} 
  P.~Romatschke and U.~Romatschke,
  Phys.\ Rev.\ Lett.\  {\bf 99}, 172301 (2007)
  [arXiv:0706.1522 [nucl-th]].

\bibitem{Staig:2011as} 
  P.~Staig and E.~Shuryak,
  J.\ Phys.\ G G {\bf 38}, 124039 (2011)
  [arXiv:1106.3243 [hep-ph]].

\bibitem{Policastro:2001yc} 
  G.~Policastro, D.~T.~Son and A.~O.~Starinets,
  Phys.\ Rev.\ Lett.\  {\bf 87}, 081601 (2001)
  [hep-th/0104066].



\bibitem{Lublinsky:2009kv} 
  M.~Lublinsky and E.~Shuryak,
  Phys.\ Rev.\ D {\bf 80}, 065026 (2009)
  [arXiv:0905.4069 [hep-ph]].

\bibitem{Lin:2008rw} 
  S.~Lin and E.~Shuryak,
  Phys.\ Rev.\ D {\bf 78}, 125018 (2008)
  [arXiv:0808.0910 [hep-th]].

\bibitem{Chesler:2010bi} 
  P.~M.~Chesler and L.~G.~Yaffe,
  Phys.\ Rev.\ Lett.\  {\bf 106}, 021601 (2011)
  [arXiv:1011.3562 [hep-th]].

\bibitem{janik2}
  M.~P.~Heller, R.~A.~Janik and P.~Witaszczyk,
  arXiv:1103.3452 [hep-th].

\bibitem{Lublinsky:2011cw} 
  M.~Lublinsky and E.~Shuryak,
  Phys.\ Rev.\ C {\bf 84}, 061901 (2011)
  [arXiv:1108.3972 [hep-ph]].

\bibitem{Hubeny:2011hd} 
  V.~E.~Hubeny, S.~Minwalla and M.~Rangamani,
  arXiv:1107.5780 [hep-th].

\bibitem{1004.3803} 
  S.~Khlebnikov, M.~Kruczenski and G.~Michalogiorgakis,
  Phys.\ Rev.\ D {\bf 82}, 125003 (2010)
  [arXiv:1004.3803 [hep-th]].
  
  \bibitem{1105.1355}
 S.~Khlebnikov, M.~Kruczenski and G.~Michalogiorgakis,
  JHEP {\bf 1107}, 097 (2011)
  [arXiv:1105.1355 [hep-th]].
  
\bibitem{CasalderreySolana:2004qm} 
  J.~Casalderrey-Solana, E.~V.~Shuryak and D.~Teaney,
  J.\ Phys.\ Conf.\ Ser.\  {\bf 27}, 22 (2005)
  [Nucl.\ Phys.\ A {\bf 774}, 577 (2006)]
  [hep-ph/0411315].
  
\bibitem{Chesler:2007sv} 
  P.~M.~Chesler and L.~G.~Yaffe,
  Phys.\ Rev.\ D {\bf 78}, 045013 (2008)
  [arXiv:0712.0050 [hep-th]].
  
\bibitem{comment} This useful comment has been made by J.Maldacena, in a conversation.

\bibitem{comment2} At discussion session at KITP program I suggested the strong shock problem as 
one of interesting physics applications of numerical relativity.  I learned later that
Jorge E. Santos (UCSB) had started working on relaxing all the Einstein equations on a grid:
at the moment of this writing no preprints/results of it are known to me, perhaps his work is
in preparation.


\end{thebibliography}
\end{document}